# Des itinéraires vers un bâtiment ou une salle adaptés aux spécificités des usagers

## L'approche adoptée dans l'application mobile OPALE

Stéphanie Jean-Daubias, Thierry Excoffier, Otman Azziz
Université Lyon 1, CNRS, INSA Lyon, LIRIS, UMR5205, F-69622 Villeurbanne, France

*Résumé*—OPALE est une application mobile multi-services à destination des usagers des campus de l'Université Claude Bernard Lyon 1. Elle associe découverte des ressources pratiques, culturelles et scientifiques à des fonctionnalités utiles au quotidien. Les points d'intérêts sont géolocalisés sur une carte du campus et atteignables via un itinéraire. Mais un itinéraire vers un bâtiment ne suffit pas toujours à se repérer. Le problème d'orientation se situe en effet souvent dans les derniers mètres : entre l'entrée du bâtiment et la salle recherchée. Dans cet article nous présentons OPALE, ainsi que l'approche que nous avons adoptée pour résoudre ce problème d'orientation des usagers dans un bâtiment. Nous montrerons également comment nous prenons en compte les spécificités des usagers, tout particulièrement ceux porteurs de handicaps, qu'ils soient physiques ou cognitifs.

*Mots-clés—application mobile, itinéraire intérieur, plus court chemin, adaptation, PMR, Troubles du Spectre Autistique*

## I. Introduction

OPALE (Outil Permettant aux Acteurs des campus de Lyon 1 d'appréhender leur Environnement pratique culturel et scientifique) [19] est une application mobile multi-services à destination des usagers des campus de l'Université Claude Bernard Lyon 1 (UCBL). Elle associe services pratiques (notes, emploi du temps, mail, horaires de transports en commun…) et outils de découverte des campus (bâtiments, équipements et services, œuvres d'art, informations scientifiques et sportives…). Un des points forts d'OPALE est la proposition de cartes des campus avec localisation des ressources utiles, en particulier les bâtiments, géolocalisation de l'utilisateur et calcul d'itinéraire vers un bâtiment donné.

Dans cet article, nous présentons l'application OPALE et ses principales possibilités, en détaillant sa fonctionnalité d'itinéraires en extérieur et en intérieur. Notre approche s'appuie sur la modélisation de bâtiments complexes prenant en compte les salles, mais aussi les droits d'accès, les étages, ailes, escaliers, ascenseurs, entrées (parfois à différents niveaux), et s'appuie sur la cartographie contributive Open StreetMap.

### A. Genèse de l'application OPALE et de ce projet

L'idée qui a conduit à la création d'OPALE a émergé en juillet 2015 lors d'une réunion sur la médiation scientifique à l'université. Ses acteurs étaient frustrés de pas réussir à valoriser le riche patrimoine culturel et scientifique du campus LyonTech-la Doua [17]. Nous avons pensé qu'une application mobile géolocalisant les éléments de ce patrimoine pourrait répondre à ce besoin. OPALE est née l'année suivante, elle s'est étoffée et ses objectifs ont été étendus pour devenir progressivement l'application telle qu'elle est présentée ici. Quant aux itinéraires vers une salle, la question a émergé de la rencontre de la responsable du projet OPALE avec une équipe d'étudiants du parcours « Pharmacien-ingénieur » de l'UCBL qui projetaient de créer une modélisation interactive 3D de Rockefeller, le très complexe bâtiment où ils ont cours. L'idée a évolué jusqu'à la solution présentée dans cet article.

### B. Objectifs de l'application et public cible

Au-delà de la motivation initiale de valorisation du patrimoine, l'objectif d'OPALE est désormais de fournir aux usagers des 13 sites et campus de l'UCBL tous les services dont ils ont besoin pour leur vie universitaire au sens large (scolarité, déplacements, restauration, culture, sport, santé…).

### C. Public cible

Si l'application est centrée sur les 45 000 étudiants de l'UCBL, elle peut également être utile à ses 5 000 personnels (dont 2 800 enseignants-chercheurs).

Dans la suite de cet article, nous présentons dans un premier temps globalement l'application OPALE, avant d'aborder le calcul d'itinéraires, d'abord vers un bâtiment, puis vers une salle. Nous terminons par une conclusion et des perspectives.

## II. L'application mobile OPALE

### A. La carte, les campus, les types de ressources, les bâtiments

La première version d'OPALE comportait uniquement la carte du campus LyonTech-la Doua à Villeurbanne. OPALE propose désormais la carte des 13 sites et campus de l'UCBL.

La carte d'un campus (cf. 1$^{er}$ écran de la Fig. 1) affiche les points d'intérêts de 29 types organisés en 5 catégories (pratique, administratif, transport, culture et sciences). On y trouve bâtiments, amphithéâtres, équipements sportifs, scolarités (avec leurs coordonnées et leurs horaires d'ouverture), services universitaires (service de santé universitaire, mission handicap, service d'orientation, etc.). Le grand nombre de points d'intérêt (plus de 1000 pour les 13 campus) nous a conduit à mettre en place un système de filtres pour permettre aux utilisateurs de choisir les types de ressources qu'ils affichent sur la carte.

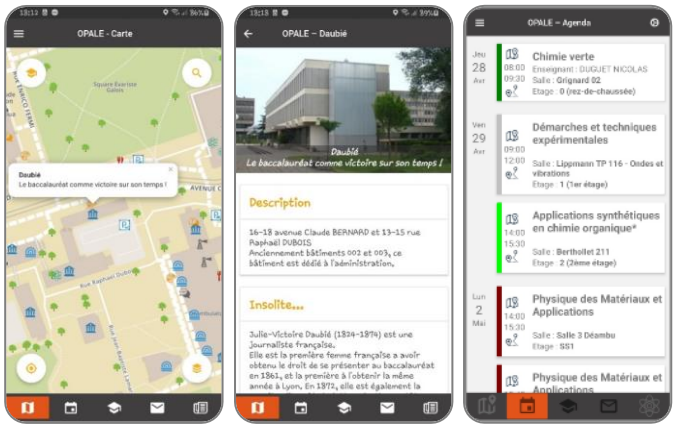

Fig. 1. Captures d'écran d'OPALE : carte, fiche d'un bâtiment, agenda.

Les informations de certains points d'intérêt sont enrichies grâce à l'utilisation de données ouvertes. Ainsi, les lieux de restauration intègrent un affichage des menus quotidiens pour les restaurants du CROUS [10], en temps réel les arrêts de tramway, de bus et métros comportent les horaires et les stations de vélo en libre-service indiquent le nombre de vélo'v disponibles [11]. Toujours concernant les transports, la carte d'OPALE affiche également les places de stationnement (voiture PMR et vélos). L'idée est de mettre en avant les approches durables et éthiques, raison pour laquelle on trouve aussi dans OPALE les défibrillateurs et des actions en faveur du développement durable (par exemple points de recyclage des déchets et hôtels à insectes).

En lien avec l'objectif initial de l'application de mettre en avant les ressources patrimoniales et scientifiques, les œuvres d'art du campus de la Doua sont également présentes, ainsi que sa collection d'arbres remarquables [21]. Chaque ressource est présentée dans une fiche (cf. 2ème écran de la Fig. 1) qui comporte une photo de ce point d'intérêt, une description textuelle factuelle, ainsi qu'un texte « insolite » abordant cette ressource de manière décalée (ça peut être des informations sur l'artiste pour les œuvres d'art, sur le scientifique qui a donné son nom au bâtiment, ou une anecdote scientifique voire littéraire en lien avec la ressource, l'idée étant de donner à réfléchir ou à sourire).

### B. Les fonctionnalités (pratiques) hors carte

L'application a rapidement été enrichie de fonctionnalités pratiques, répondant ainsi aux attentes des étudiants, ainsi qu'à une demande institutionnelle.

Une partie de ces outils sont accessibles après connexion au système d'information de l'université via le CAS (Central Authentication Service), un système d'authentification unifié. Les étudiants ont alors accès à leurs notes (via l'outil TOMUSS [23]), à leur agenda individuel (en lien avec ADE, cf. 3ème écran de la Fig. 1), aux environnements numériques de travail (Claroline Connect [7] et Moodle), ainsi qu'à leur mail universitaire, sans qu'aucune configuration ne soit nécessaire. Ils ont également accès aux informations concernant la bibliothèque universitaire (réservation de salles de travail, recherche de documents numériques et physiques, livres à rendre, messagerie interne), au suivi des conventions de stage, au service d'orientation, à l'annuaire, au site Web de l'UCBL, au portail Étudiants, à la mission égalité-diversité, la mission handicap, etc. Une fiche SOS rassemble les numéros de secours du campus et nationaux, les contacts santé, un lien vers les petites annonces, les infos pour se connecter au Wifi, recharger sa carte Izly, reporter un problème technique, etc.

L'appli propose aussi des actualités scientifiques (Sciences pour tous [22], Le journal du CNRS [18]) et sportives.

### III. DES ITINÉRAIRES POUR OPALE

OPALE et sa carte ont rapidement été utilisés davantage pour la localisation des bâtiments que pour l'inventaire des œuvres d'art et des arbres remarquables qui était à l'origine de sa création. L'idée de proposer des itinéraires pour atteindre les bâtiments a ainsi émergé, notamment en raison de l'étendue du campus de la Doua (100 hectares) et du nombre important de bâtiments qu'il comporte (plus de 100).

#### A. Des itinéraires Google Maps vers un bâtiment

Dans un premier temps, une solution simple a été mise en place pour fournir aux usagers de l'appli le tracé sur la carte des itinéraires entre leur position géolocalisée et le bâtiment recherché. Elle s'appuyait sur un fond de carte Google Maps et l'API associée pour le calcul d'itinéraires [3], couplés à un module de recherche des bâtiments. Sur cette base, OPALE fournissait un itinéraire au choix pour piétons, cyclistes ou automobilistes. Les itinéraires obtenus étaient toutefois peu fiables à l'intérieur du campus, en particulier pour les piétons. En effet, la cartographie Google Maps du campus est assez peu détaillée et pas toujours fiable.

Pour cette raison couplée à des raisons de gratuité de service, nous avons ensuite opté pour la cartographie collaborative Open StreetMap (OSM) [20] couplée à des API de calcul d'itinéraires [4]. L'ensemble fournit les mêmes services que la configuration précédente, mais avec de meilleures performances pour les itinéraires piétons, avec en outre la possibilité pour nous d'améliorer rapidement la cartographie en cas de faiblesse identifiée. En effet, moyennant la création d'un compte, il est possible de contribuer finement aux cartes de façon adaptée à nos besoins, mais surtout de voir nos ajouts opérationnels immédiatement. Pour cela nous pouvons de plus nous appuyer sur une communauté de contributeurs, en particulier cyclistes, active sur le campus.

En lien avec ces itinéraires, une nouvelle fonctionnalité a été introduite : l'accès à un itinéraire à partir de l'agenda de l'étudiant vers le bâtiment où il a cours (cf. icône au-dessous de l'horaire du cours sur la capture d'écran au centre de la Fig. 1).

#### B. Des itinéraires OPALE vers une salle

##### 1) Le problème des 100 derniers mètres

Dans un second temps, nous avons progressivement identifié un besoin d'orientation complémentaire : ce que nous appelons le « problème des 100 derniers mètres ». Cette expression est un clin d'œil au problème du dernier kilomètre, celui qui pose actuellement le plus de défis logistiques pour les livraisons [9]. En effet, proposer un itinéraire vers un bâtiment est une solution satisfaisante dans la plupart des cas pour les premières visites de bâtiments simples, mais dans le cas de bâtiments complexes, la difficulté au quotidien est plus de s'orienter dans le bâtiment que de trouver le bâtiment lui-même. Nous pensons que les 100 derniers mètres sont un défi majeur pour l'orientation des usagers, d'autant plus s'ils sont porteurs de handicaps, qu'ils soient physiques (PMR) ou cognitifs, notamment TSA (Troubles du Spectre Autistique). En lien avec les étudiants du parcours « Pharmacien-ingénieur » et avec la mission handicap de l'UCBL, nous avons ainsi défini un projet de cartographie intérieure de bâtiments couplée à une extension des itinéraires (non plus seulement vers un bâtiment mais également vers une salle), tout en cherchant à intégrer la notion d'adaptation des itinéraires à des besoins spécifiques (prenant en compte des handicaps physiques et cognitifs).

La problématique du travail de recherche qui en a découlé peut se formuler ainsi : comment guider un utilisateur souhaitant aller d'un point A (en extérieur ou en intérieur) à un point B (en extérieur ou en intérieur) et prenant en compte ses besoins spécifiques de déplacement ?

Ce travail a nécessité la définition d'un modèle des différents constituants d'un plan intérieur, ainsi qu'un modèle des contraintes et préférences de l'utilisateur compatible avec les handicaps. Il a également conduit à mettre en œuvre un algorithme calculant un itinéraire adapté au contexte et aux spécificités de l'utilisateur, et enfin à proposer une interface utilisateurs permettant d'afficher sur le plan intérieur d'un bâtiment l'itinéraire calculé vers la salle visée.

*2) Les systèmes de cartographie*

Comme nous l'avons vu dans la section III.A, nous avions préalablement comparé les services existants de cartographie et de calcul d'itinéraires associés avant de porter notre choix sur Open StreetMap. Afin d'étendre la cartographie d'OPALE à l'intérieur des bâtiments, nous avons refait cette étude en nous concentrant cette fois sur les services offrant cette possibilité. À cette époque (2022), Google Maps et OSM commençaient tout juste à permettre d'établir et d'exploiter des plans en intérieur, mais pour l'un comme pour l'autre, cette possibilité était très peu utilisée. Contrairement à celui de Google Maps, dans le service d'OSM l'ajout de données, la récupération et l'affichage de fonds de cartes est entièrement libre et gratuit. Ajoutés à la rapidité du processus de mise en ligne et à l'aspect contributif de l'approche, ces avantages nous ont conduits à choisir OSM.

*3) Le calcul d'itinéraire et la modélisation d'un bâtiment*

Établir **un itinéraire** entre deux points est un problème connu sous le nom de problème du plus court chemin et résolu depuis de nombreuses années. Il s'agit dans un graphe pondéré, orienté ou non, de proposer un chemin de coût minimal entre deux sommets, c'est-à-dire un chemin dont la somme des poids des arcs ou arêtes est minimale. Notre besoin d'itinéraire peut être modélisé par un graphe pondéré non orienté où les points d'intérêts sont représentés par les sommets du graphe et les cheminements entre eux par les arêtes du graphe. Les arêtes sont pondérées par le coût de déplacement (classiquement la distance) entre les deux points d'intérêt qu'elles relient.

Nous définissons un *lieu* (cf. Fig. 2) comme un point d'intérêt (exemple : une salle d'enseignement, un bureau, des toilettes, des escaliers, un ascenseur) intérieur à un bâtiment.

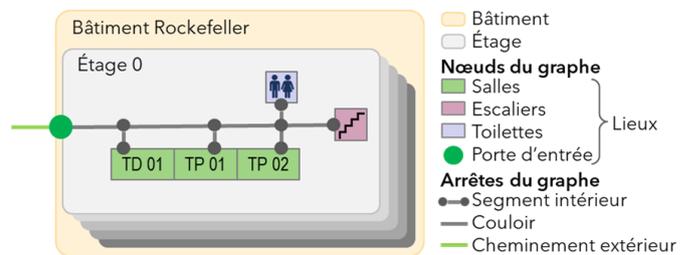

Fig. 2. Modélisation du graphe : lien entre les lieux par des segments.

Pour être atteignable dans un itinéraire, un lieu doit être relié aux cheminements du bâtiment (généralement à un couloir) par un segment. Un lieu est modélisé comme un nœud du graphe. Un *segment* est défini comme la connexion entre deux nœuds, il est modélisé comme une arête du graphe. Un cheminement entre deux lieux est donc un ensemble de segments reliés entre eux. À un segment est associé un poids, qui représente la distance entre les deux nœuds. Dans le calcul de distance, notre algorithme prend en compte à la fois la distance cumulée des segments concernés, mais aussi le nombre de nœuds parcourus. En effet, les nœuds correspondent généralement à des changements de direction, qui, s'ils n'augmentent pas la distance entre deux points, augmentent en revanche le temps pour parcourir le cheminement entre deux lieux.

Plusieurs algorithmes du plus court chemin sont disponibles. Les algorithmes « All-pairs » comme celui de Floyd-Warshall [12], cherchent les plus courts chemins entre tous les couples de sommets du graphe, ce qui est très coûteux et va au-delà de nos besoins. Les algorithmes « Single-source » calculent les plus courts chemins entre un sommet donné et tous les autres sommets du graphe, simplifiant ainsi le problème en envisageant le graphe comme un arbre, avec des poids quelconques (Bellman-Ford [5]) ou positifs (Dijkstra [8]). Certains algorithmes sont rendus plus efficaces par l'utilisation d'heuristiques (A* [13]) pour optimiser un chemin entre deux sommets donnés. N'ayant pas identifié heuristiques pertinentes, nous n'avons pas opté pour cette approche. L'approche de Dijkstra est celle qui semble la plus adaptée à notre problème. Dans notre première mise en œuvre, l'itinéraire était calculé grâce à un algorithme du plus court chemin entre deux points, décliné à partir de celui de Dijkstra.

*4) La personnalisation des itinéraires*

L'objectif de notre projet n'était pas seulement de proposer l'itinéraire le plus court, mais aussi et surtout l'itinéraire le plus adapté aux besoins de l'usager, c'est-à-dire le chemin le plus court respectant les contraintes exprimées par l'utilisateur.

Pour cela nous avons identifié puis modélisé les contraintes qui nous semblaient pertinentes dans nos contextes d'usages présents et à venir d'OPALE, en nous centrant sur les usagers porteurs de handicaps, mais en adoptant une approche plus large que nous voulons inclusive et non stigmatisante.

Nous n'avons pas trouvé d'algorithme du plus court chemin vraiment adapté à cette approche [1]. En complément de notre définition des lieux et segments, nous définissons donc une *caractéristique* comme une propriété des lieux et des segments pouvant impacter l'itinéraire proposé (par exemple marquages podotactiles, portes lourdes). Et nous définissons un *critère de déplacement* comme l'importance pour l'utilisateur (en positif ou en négatif) d'une caractéristique, il est représenté par une valeur numérique exploitée comme poids dans notre algorithme du plus court chemin.

| Critère | Importance | Nom | Facteur |
|---|---|---|---|
| Contrainte | Positive | Indispensable pour moi | 1000 |
| Préférence | Positive | Je veux | 100 |
| Préférence | Positive | Je préfère | 10 |
| Neutre | Neutre | Neutre | 1 |
| Préférence | Négative | Je ne préfère pas | 10 |
| Préférence | Négative | Je ne veux pas | 100 |
| Contrainte | Négative | Impossible pour moi | 1000 |

Fig. 3. Critères de personnalisation et facteurs associés.

Nous avons modélisé dix critères de personnalisation couplés à sept niveaux de valuation, couvrant à la fois des handicaps physiques et sensoriels, mais aussi des préférences indépendantes de tout handicap. Nos critères de personnalisation sont : ascenseurs, escaliers, lieu calme, zone éclairée, marquage podotactile, porte automatique, porte lourde, rampe, terrain difficile et zone de travaux (cf. partie centrale de la Fig. 7). Les valeurs attribuables sont rédigées du point de vue de l'utilisateur et non du système : par exemple Indispensable pour moi, Je veux, Je préfère (cf. Fig. 7). Certaines valeurs font des critères des contraintes (Indispensable, Impossible), d'autres en font des préférences (Je veux, Je préfère). L'objectif est que tout utilisateur puisse se sentir concerné par l'écran de paramétrage des itinéraires d'OPALE afin que cette fonctionnalité soit mieux connue et non stigmatisante. En effet, un étudiant en fauteuil roulant pourra configurer Escaliers / Impossible, Ascenseurs / Indispensable, Porte lourde / Je ne veux pas ; un enseignant qui souhaite faire plus d'exercice pourra choisir Escaliers / Je préfère, Ascenseurs / Je ne préfère pas ; un étudiant autiste pourra par exemple sélectionner Lieu calme / Je veux et Terrain difficile / Je ne préfère pas.

Dans le calcul du plus court chemin personnalisé, notre algorithme pondère les poids existants du graphe par les facteurs donnés dans la Fig. 3. Si un nœud (ex : une porte) ou un segment (ex : un couloir) est associé à des caractéristiques prises en compte dans la personnalisation (ex : une porte lourde, un couloir bruyant), alors l'algorithme prend en compte les sources d'efforts pour l'utilisateur : si l'importance de la caractéristique est positive dans les paramètres de l'utilisateur, le poids du segment sera divisé par le facteur donné par la Fig. 3, augmentant ainsi les chances que ce cheminement soit utilisé dans l'itinéraire ; si l'importance est négative, le poids sera multiplié par ce facteur, diminuant ainsi le risque que ce cheminement soit proposé à l'utilisateur.

### 5) Le calcul d'itinéraires alternatifs

Nous sommes conscients que les itinéraires proposés ne seront pas forcément aussi bien adaptés à l'usager que nous l'envisagions. Nous avons donc mis en place une fonctionnalité de proposition d'itinéraires alternatifs permettant à l'utilisateur de choisir celui qui lui semble le plus adapté.

Établir **plusieurs itinéraires** alternatifs entre deux points est un problème connu sous le nom de problème des k plus courts chemins et résolu par une généralisation des algorithmes du plus court chemin [6]. Ainsi, l'algorithme de Hoffman, très efficace, s'appuie sur un premier plus court chemin pour lequel il cherche ensuite des variantes [14]. Il nécessite des adaptations pour proposer des alternatives réellement différentes du premier chemin trouvé.

Après avoir étudié plusieurs possibilités, nous avons fait le choix de simplifier les propositions faites à l'utilisateur en lui soumettant seulement deux itinéraires : l'itinéraire le plus rapide parmi ceux respectant tous ses critères au mieux, et l'itinéraire le plus rapide sans prendre en compte ses critères. L'idée est de permettre à l'utilisateur de comparer les deux et de s'appuyer sur son éventuelle expérience personnelle pour choisir d'utiliser tout ou partie du chemin le plus rapide même si en théorie il ne respecte pas ses critères. Par exemple, notre étudiant autiste pourra prendre l'itinéraire le plus court indiqué comme bruyant car il sait qu'à cette heure-là le couloir est assez calme ; notre enseignant sportif décidera de prendre exceptionnellement l'ascenseur pour arriver à l'heure en cours sans être essoufflé.

Pour cela, l'algorithme d'OPALE applique dans un premier temps son algorithme du plus court chemin tel que décrit dans la section III.B.3) pour proposer l'itinéraire le plus court entre le point de départ et le point d'arrivée sans prise en compte des paramètres de l'utilisateur. Dans un second temps, on applique les préférences de l'utilisateur selon la méthode décrite dans la section III.B.4). Les nouveaux poids ainsi calculés sont appliqués à notre graphe avant d'appliquer de nouveau notre algorithme du plus court chemin qui produit cette fois un itinéraire adapté aux spécificités de notre utilisateur si un tel itinéraire existe. En effet, sur la base de la description des cheminements existants, certains points ne sont pas atteignables en respectant certains critères : il reste par exemple malheureusement certaines salles ou bâtiments pour lesquels la mise en accessibilité pose problème et qui ne sont pas entièrement accessibles en fauteuil roulant, nous reviendrons sur ce point dans les perspectives.

### 6) Mise en œuvre logicielle

L'architecture générale d'OPALE (cf. Fig. 4) associe une interface Web qui permet à l'administrateur de gérer les contenus, couplée à un backend qui soutient le fonctionnement de l'application mobile elle-même. Les données sur les points d'intérêts d'OPALE sont stockées dans une base de données qui est exploitée par l'appli mobile.

En ce qui concerne la proposition d'itinéraires présentée dans cet article, son fonctionnement est le suivant. Nos données cartographiques sont intégrées à OSM (cf. section 7) pour la description du processus de contribution des données). En cas d'ajout ou de mise à jour des données, l'administrateur lance un moissonnage des données du campus concerné, et une vérification des données. Les nouveaux plans sont ensuite disponibles dans l'appli et sont exploités lorsque les utilisateurs demandent à consulter un plan ou à obtenir des itinéraires vers un bâtiment ou vers une salle.

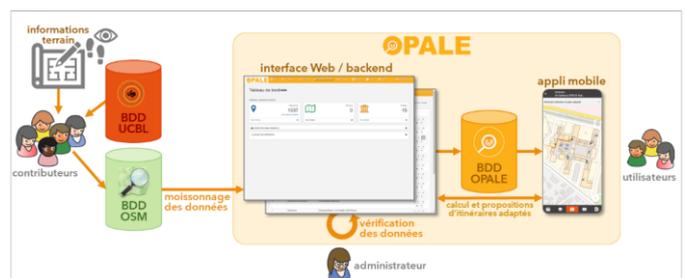

Fig. 4. Architecture d'OPALE pour les itinéraires.

*7) Ajout des données cartographiques d'intérieur*

Dans un premier temps, nous nous sommes concentrés sur le bâtiment Rockefeller du campus Lyon Santé Est de l'UCBL, de structure complexe (plusieurs étages, ailes, entrées, escaliers, ascenseurs, mais aussi des accès limités).

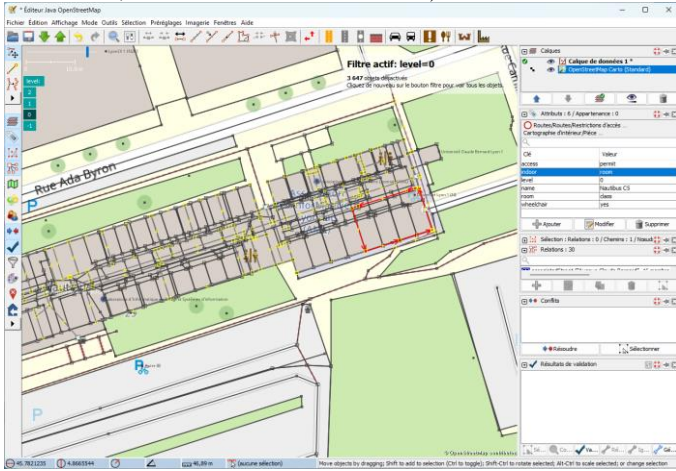

Fig. 5. Édition des propriétés d'une salle de cours avec JOSM.

L'ajout de données cartographiques en intérieur à OSM se fait soit directement dans OSM [20] [24], soit via des outils complémentaires d'édition de carte comme JOSM [16] (cf. Fig. 5). Les bâtiments sont en général déjà présents, ainsi que les cheminements extérieurs. Notre contribution cartographique consiste donc à tracer chaque salle de chaque étage et à lui attribuer les propriétés OSM adaptées (étage, nom, affectation : salle d'enseignement ou bureau par exemple, accessibilité, etc.), puis à tracer les cheminements permettant d'atteindre ces salles, les y relier et leur attribuer les propriétés adaptées (principalement concernant l'accessibilité), sans oublier de relier les cheminements intérieurs aux cheminements extérieurs existants via les portes d'entrée du bâtiment.

Ce travail nécessite des connaissances précises sur le bâtiment à cartographier. Ainsi, nous nous appuyons sur la cartographie fournie par la direction du patrimoine, sur les informations mises à disposition par les services informatiques, mais surtout sur des données récoltées sur le terrain. Pour le bâtiment Rockefeller, l'équipe d'étudiants « pharmaciens-ingénieurs » étaient particulièrement bien placés pour établir une cartographie de qualité, en lien avec l'étudiant développeur.

*8) Validation des données cartographiques*

Les données contribuées sur OSM, puis moissonnées par OPALE font ensuite l'objet d'une validation soigneuse, afin notamment de repérer les inévitables erreurs de cartographie, en particulier dans les cheminements et les accès aux salles.

Pour cela, un algorithme calcule pour un bâtiment donné tous les itinéraires possibles entre chaque salle pour garantir la validité des données contribuées sur OSM et l'existence d'itinéraires pour chaque couple de lieux de départ / d'arrivée possible et pour chaque configuration de critères de déplacement possible. Pour des raisons de temps de calcul, c'est une version simplifiée, sans ce dernier point, qui a été mise en œuvre. Elle indique le taux de couples de deux lieux d'un bâtiment pour lesquels il est impossible de calculer un itinéraire. Tant que ce taux n'est pas nul, il existe au moins un lieu non ou mal relié aux cheminements du bâtiment. La cartographie doit donc être corrigée ou complétée.

*9) Exploitation de la cartographie et des itinéraires*

La nouvelle version d'OPALE intégrant cette fonctionnalité permet donc non seulement d'afficher des plans détaillés des différents étages des bâtiments cartographiés (cf. 1$^{er}$ écran de la Fig. 7), mais aussi de fournir des itinéraires depuis la position de l'utilisateur ou une salle donnée vers un bâtiment ou une salle (en passant ou pas par l'extérieur, un itinéraire pouvant commencer dans une salle, indiquer comment sortir du bâtiment, atteindre le bâtiment cible puis la salle recherchée, cf. 3$^{ème}$ écran de la Fig. 7). Notons que l'appli ne gère pour l'instant pas les déplacements de l'utilisateur dans le bâtiment pendant son parcours pour des raisons pragmatiques. En effet, la géolocalisation par GPS en intérieur n'est pas performante et les alternatives actuelles (Beacons Bluetooth, balises RFID, Wi-Fi) nous semblent peu adaptées à cette phase exploratoire de notre projet. En revanche, les itinéraires fournis prennent en compte les paramètres définis par l'utilisateur. Dans un premier temps, seuls les paramètres Ascenseur et Escaliers sont opérationnels, notamment faute de contribution dans la cartographie pour les autres paramètres (par exemple zone de travaux, lieu calme).

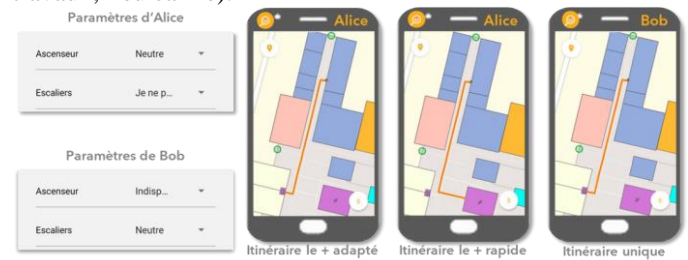

Fig. 6. Exemple de personnalisation d'itinéraires selon les paramètres.

Ainsi, selon les paramètres choisis par l'utilisateur, les itinéraires pourront différer. Prenons l'exemple d'Alice et Bob, étudiants de Pharmacie voulant aller du local de leur association situé au RDC du bâtiment Rockefeller à la BU située au 2$^{ème}$ étage du même bâtiment (cf. Fig. 6). Alice a indiqué dans son profil qu'elle ne peut pas prendre l'ascenseur (facteur 100) et a laissé la valeur neutre pour les escaliers (facteur 1). Quant à Bob, il a indiqué que les ascenseurs étaient indispensables pour lui (facteur 1000), les autres paramètres étant laissés avec une valeur neutre. Alice se verra proposer l'itinéraire le plus adapté prenant en compte ses paramètres, un itinéraire empruntant l'ascenseur, mais elle pourra aussi voir un itinéraire plus rapide empruntant l'escalier, non compatible avec ses préférences. Quant à Bob, on lui proposera dans ce cas un itinéraire unique utilisant l'ascenseur car la valeur qu'il a indiquée pour le paramètre escaliers empêche tout usage d'un tel cheminement.

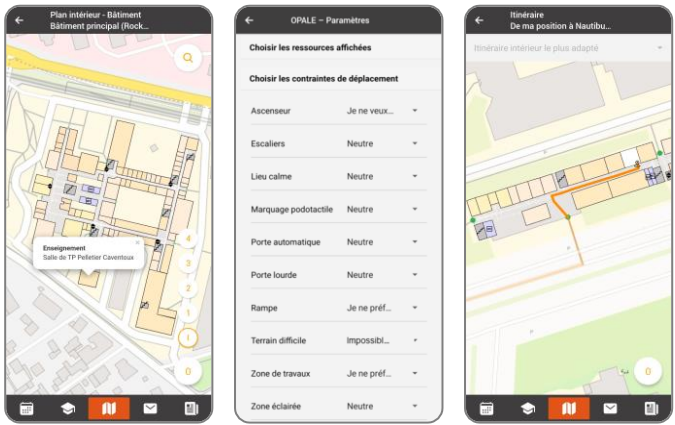

Fig. 7. Captures d'écran d'OPALE : plan intérieur, paramètres et itinéraire.

## IV. CONCLUSION ET PERSPECTIVES

Dans cet article, nous avons présenté l'application OPALE qui propose aux usagers des campus de l'Université Lyon 1 des outils pratiques au service de leur quotidien. Plus de 1000 ressources d'une trentaine de types sont disponibles dans l'appli, réparties entre les 13 sites et campus de notre université. Nous avons également présenté la fonctionnalité que nous avons récemment ajoutée pour proposer des itinéraires personnalisés entre la localisation de l'utilisateur et le bâtiment ou la salle où il cherche à se rendre. Cette fonctionnalité s'appuie d'une part sur un algorithme du plus court chemin adapté de celui de Dijkstra couplé à une gestion de paramètres et d'autre part sur une cartographie contributive utilisant Open StreetMap. Elle est disponible pour le bâtiment Rockefeller du campus Lyon Santé Est et cinq bâtiments du campus LyonTech-la Doua. Pour l'instant seuls deux paramètres sont pris en compte mais les paramètres ont été réfléchis plus largement. L'application et cette fonctionnalité ont reçu un bon accueil de la part des étudiants. Le nombre de téléchargements de l'appli croit régulièrement (2021 : 4864, 2022 : 8763, 2023 : 10061).

De nombreuses perspectives sont envisagées pour OPALE, en particulier concernant les itinéraires au centre de cet article.

Une première perspective concerne l'élargissement des paramètres pris en compte, ce qui nécessite d'une part une évolution de l'appli et de l'algorithme du plus court chemin, et d'autre part d'informations fiables et suffisamment nombreuses associées aux cartes. Nous allons prochainement travailler sur ces deux axes : à travers une réflexion plus poussée sur les paramètres pertinents pour les usagers porteurs des troubles du spectre autistique (TSA) et par un travail sur la contribution aux informations. En effet, nous avons envisagé dès le début de ce projet de mettre en place une interface permettant à des contributeurs extérieurs au projet d'alimenter la cartographie avec des informations issues du terrain. En particulier, la notion de zone calme/bruyante, pertinente pour des personnes porteuses autisme, est difficile à établir par des personnes non concernées ou non averties, en revanche, des étudiants autistes sont à même de donner cette information et pourrait être enclins à le faire. Nous allons travailler pour cela avec le INCLUDE [15], la mission handicap de Lyon 1, en lien avec Atypie Friendly [2] avec laquelle l'UCBL a signé un partenariat.

Une seconde perspective consisterait à étendre cette interface contributeurs pour recueillir des informations plus pertinentes sur des aspects plus variés des salles et des cheminements. Ces informations pourraient d'une part permettre d'améliorer encore les itinéraires proposés aux utilisateurs d'OPALE, et d'autre part être exploitées pour faire l'inventaire des problèmes et favoriser leur résolution. Par ailleurs, l'algorithme de vérification des cheminements pourrait être exploité pour identifier les difficultés d'accessibilité de certaines salles, couloirs ou bâtiments, en lien avec la direction du patrimoine, afin d'identifier les travaux d'accessibilité nécessaires.

Enfin, nous souhaitons déployer OPALE dans d'autres universités, à commencer Toulouse III via Atypie Friendly.

## V. REMERCIEMENTS

## VI. RÉFÉRENCES